\documentclass{aa}
\usepackage{natbib}
\usepackage{graphicx}
\usepackage{times}
\bibpunct{(}{)}{;}{a}{}{,}
\newcommand{\yc}{\hbox{Y~Cyg }}
\newcommand{\ye}{\hbox{Y~Cyg}}

\newcommand{\phoebe}{\hbox{\tt PHOEBE }}
\newcommand{\phoebee}{\hbox{\tt PHOEBE}}
\newcommand{\fotel}{\hbox{\tt FOTEL }}
\newcommand{\fotele}{\hbox{\tt FOTEL}}
\newcommand{\korel}{\hbox{\tt KOREL }}
\newcommand{\korele}{\hbox{\tt KOREL}}
\newcommand{\spefo}{\hbox{\tt SPEFO }}

\newcommand{\iraf}{\hbox{\tt IRAF }}
\newcommand{\AAm}{\hbox{\AA\,mm$^{-1}$}}

\newcommand{\z}{\hbox{\v{z}}}

\newcommand{\dl}{\hbox{$\bigtriangleup$}}
\newcommand{\ubv}{\hbox{$U\!B{}V$}}
\newcommand{\bv}{\hbox{$B\!-\!V$}}
\newcommand{\ub}{\hbox{$U\!-\!B$}}

\newcommand{\oc}{\hbox{$O\!-\!C$}}

\newcommand{\p}{$\pm$}

\newcommand{\m}{$^{\rm m}\!\!.$}
\newcommand{\D}{$^{\rm d}\!\!.$}

\newcommand{\kms}{km~s$^{-1}$ }
\newcommand{\ks}{km~s$^{-1}$}
\newcommand{\vsin}{$v$~sin~$i$ }

\newcommand{\tef}{$T_{\rm eff}$ }
\newcommand{\lgg}{{\rm log}~$g$ }
\newcommand{\ms}{M$_{\odot}$}
\newcommand{\rs}{R$_{\odot}$}

\newcommand{\ha}{H$\alpha$ }

\begin{document}

    \title{Revised physical elements of the astrophysically important
           O9.5+O9.5V eclipsing binary system \ye
\thanks{Based on new spectral and photometric observations from
 the following observatories: Dominion Astrophysical Observatory,
 Hvar, Ond\v{r}ejov, Fairborn, and Sejong}
\thanks{Tables 4 and 5 are available only in electronic form
 at the CDS via anonymous ftp to cdarc.u-strasbg.fr (130.79.128.5)
 or via http://cdsweb.u-strasbg.fr/cgi-bin/qcat?J/A+A/}
}
\author{P. Harmanec\inst{1}\and D. E. Holmgren\inst{2}\and M.~Wolf\inst{1}
\and H. Bo\z i\'c\inst{3}\and
E.F. Guinan\inst{4}\and Y.W. Kang\inst{5}\and
P.~Mayer\inst{1}\and G.P.~McCook\inst{4}\and
J.Nemravov\'a\inst{1}\and S.~Yang\inst{6}\and M.~\v{S}lechta\inst{7}\and
D.~Ru\v{z}djak\inst{3}\and D.~Sudar\inst{3}\and P.~Svoboda\inst{8}}

   \offprints{P. Harmanec,\\
       \email Petr.Harmanec@mff.cuni.cz}

  \institute{
   Astronomical Institute of the Charles University, Faculty of
   Mathematics and Physics, \hfill\break
   V~Hole\v{s}ovi\v{c}k\'ach~2, CZ-180 00 Praha~8 - Troja, Czech Republic
\and
   SMART Technologies, 3636 Research Road N.W., Calgary, Alberta,
   T2L 1Y1 Canada
\and
   Hvar Observatory, Faculty of Geodesy, Zagreb University,
   Ka\v{c}i\'ceva 26, 10000 Zagreb, Croatia
\and
   Department of Astronomy and Astrophysics, Villanova University,
   Villanova, PA 19085, USA
\and
   Dept. of Earth Sciences, Sejong University Seoul, 143-747, Korea
\and
   Physics \& Astronomy Department, University of Victoria,
   PO Box 3055 STN CSC, Victoria, BC, V8W 3P6, Canada
\and
   Astronomical Institute, Academy of Sciences of the Czech Republic,
   251 65 Ond\v{r}ejov, Czech Republic
\and
   TRIBASE Net, ltd., V\'ypustky 5, CZ-614 00 Brno, Czech Republic
}

\date{Received \today}


   \date{Release \today}

  \abstract
   {Rapid advancements in light-curve and radial-velocity curve
modelling, as well as improvements in the accuracy of observations, allow
more stringent tests of the theory of stellar evolution.
Binaries with rapid apsidal advance are particularly useful in this
respect since the internal structure of the stars can also be tested.}
   {Thanks to its long and rich observational history and rapid apsidal
motion, the massive eclipsing binary \yc represents one of the cornerstones
of critical tests of stellar evolutionary theory for massive stars.
Nevertheless, the determination
of the basic physical properties is less accurate than it could be given
the existing number of spectral and photometric observations. Our goal is
to analyse all these data simultaneously with the new dedicated series
of our own spectral and photometric observations from observatories widely
separated in longitude.}
   {We obtained new series of \ubv\ observations at three observatories
separated in local time to obtain complete light curves of \yc for its
orbital period close to 3 days. This new photometry was reduced and
carefully transformed to the standard \ubv\ system using the HEC22 program.
We also obtained new series of red spectra secured at two observatories
and re-analysed earlier obtained blue electronic spectra.
Reduction of the new spectra was carried out in the \iraf and \spefo
programs. Orbital elements were derived independently with the \fotel and
\phoebe programs and via disentangling with the program \korele. The final
combined solution was obtained with the program \phoebe.}
{Our analyses provide the most accurate value of the apsidal period of
($47.805\pm0.030$)~yrs published so far and the following physical elements:
$M_1=17.72\pm0.35$~\ms, $M_2=17.73\pm0.30$~\ms, $R_1=5.785\pm0.091$~\rs,
and $R_2=5.816\pm0.063$~\rs.
The disentangling thus resulted in the masses,
which are somewhat higher than all previous determinations and virtually the
same for both stars, while the light curve implies a slighly higher radius
and luminosity for star~2.
The above empirical values imply the logarithm of the internal
structure constant $\log~k_2 = -1.937.$ A comparison with Claret's stellar
interior models implies an age close to $2\times10^6$~yrs for both stars.
}
   {The claimed accuracy of modern element determination of 1--2 per cent
still seems a bit too optimistic and obtaining new high-dispersion and
high-resolution spectra is desirable.}

   \titlerunning{Revised physical elements of \ye}
   \authorrunning{Harmanec et al.}

   \maketitle
%

\section {Introduction}
The initial motivation of this study was to search for possible
line-profile variability in early--type binary systems, and
to derive new orbital and physical properties of stars in
eclipsing binaries. This project (known as SEFONO)
has been discussed in detail in the first two papers of this series devoted
to V436~Per and $\beta$~Sco (see \citealt{sef1} and \citealt{sef2}).
This paper is devoted to a detailed study
of the astrophysically important high-mass eclipsing system \ye.
While compiling and analyzing the rich series of its observations,
we realised that for such a well-studied system it is valuable to
investigate how sensitive the determination
of its basic physical properties is to the various assumptions about
the input physics and also to the methods of analysis used.

Y~Cygni (HD~198846, BD+34$^\circ$4184) is a very interesting system from
the point of view of the theory of stellar structure.
It is a double-lined spectroscopic and eclipsing binary with an
eccentric orbit of 2.996~d and a relatively rapid apsidal motion
($P_{\rm aps.}$ = 48~yrs). \yc was among the very first binaries
for which the apsidal motion was convincingly detected. Thanks to this,
it became the subject of numerous studies. The history of its
investigation is summarised in the papers by \citet{hill95} and
\citet{hol95} and need not be repeated here.
We only mention some more recent studies
relevant to the topic. \citet{simon92,simon94} developed a new
technique of spectra disentangling, which permits the separation of the
spectra of binary components with a simultaneous determination of the orbital
and physical elements. \citet{simon94b} applied this technique to a study of \ye.
Comparing the disentangled and synthetic spectra, they obtained a new
estimate of the \tef and \lgg for both components.  New orbital elements of
\yc  were derived from the published radial velocities (RVs)
of \citet{hill95} by \citet{kar2007}, who used a least-squares technique,
and by \citet{kar2009}, who used the artificial neural network method
and who arrived at virtually identical masses for both stars.

Several authors argued in favour of the presence of stellar wind in \ye.
\citet{moris84} studied and modelled the resonance \ion{C}{IV} and
\ion{N}{V} UV lines in the spectra from the
International Ultraviolet Explorer (IUE).
They found the lines are blue-shifted for some 160 \kms with respect
to photospheric lines, derived a terminal velocity of 1500 \ks, and
estimated a rather moderate mass-loss rate
of $10^{-8.4}$~\ms\,yr$^{-1}$. \citet{koch89} studied polarimetric changes
of \yc and concluded that there is some evidence of hot material between
the stars, although they admitted that the observed variations are not large,
and argued again in favour of the presence of stellar wind. Their $B$-band
polarimetric observations were later re-analyzed by \citet{fox94}, who
could not find any periodicity related to the binary orbit. He noted
that there was some evidence from observations close to binary eclipses
that one of the stars possesses a stellar wind. Finally, \cite{pfei94}
studied and modelled the stellar wind from the UV line profiles of
\ion{C}{IV} and \ion{Si}{IV}. To this we should add that our \ha
spectra do not show any detectable emission.

In spite of all the efforts, the basic physical properties of the system
have not been derived accurately enough. This is mainly because of
the awkward orbital period that is very close to 3 days. Another complication
is a close similarity of the spectra of both stars and a mass ratio near 1.0\,.
In Table~\ref{history} we summarize the values of the mass ratio
and individual masses derived by various investigators. Since different
investigators of \yc have identified the primary and secondary differently,
we shall not use the terms {\sl primary} and {\sl secondary}, but
will call the components {\sl star~1} and {\sl star~2}, with the convention
that the minimum RV of star~1 occurs near JD~2446308.97 for the sidereal
orbital period of 2.99633179~d (see below).

\begin{table}
\caption[]{Individual masses and the mass ratio $M_2/M_1$ derived by various
investigators. Since different authors alternatively denoted one or the other
component of the pair as the primary, we adopt the convention
that star~1 is the component that has the minimum RV near JD~2446308.97 for
the sidereal orbital period of 2.99633179~d (see the text below).
}\label{history}
\begin{flushleft}
\begin{tabular}{ccccllrl}
\hline\hline\noalign{\smallskip}
  $M_2/M_1$&$M_1\sin^3i$&$M_2\sin^3i$& Source&Note\\
   & (\ms)      & (\ms)      & \\
\noalign{\smallskip}\hline\noalign{\smallskip}
0.922      &$16.5$      &$15.2$      &A\\
1.011      &$17.2$      &$17.4$      &B\\
1.037      &$16.4\pm0.3$&$17.0\pm0.3$&C\\
1.024      &$16.6\pm0.2$&$17.0\pm0.3$&D\\
1.031      &$16.94\pm0.11$&$17.47\pm0.11$&E&*\\
0.949      &$17.6\pm0.4$&$16.7\pm0.5$&F&*\\
0.989      &$17.4\pm0.3$&$17.2\pm0.2$&G\\
1.002      &$17.1\pm0.4$&$17.14\pm0.12$&H\\
0.998      &$17.26\pm0.22$&$17.22\pm0.22$&I\\
\noalign{\smallskip}\hline\noalign{\smallskip}
\end{tabular}
\tablefoot{
{\it Abbreviations used in Col. 4:} \ \
A... \citet{plaskett20};
B... \citet{redman};
C... \citet{vitr};
D... \citet{stick09};
E... \citet{simon94b};
F... \citet{burk97};
G... \citet{hill95};
H... \citet{kar2007};
I... \citet{kar2009}.\\
{\it An asterisk in Col. 5 denotes that the role of the components
is inverted with respect to our convention in the original study.}
}
\end{flushleft}
\end{table}

Usually, the apsidal motion was investigated on the basis of existing times of
photometric minima. In spite of the obvious importance of this binary,
no really complete light curve has been obtained within one specific
epoch, and a light curve in a standard system has not been obtained yet.
This has prevented the determination of the individual colours
of the binary components, and therefore their temperatures.

\begin{table*}
\caption[]{RV data sets}\label{jourv}
\begin{flushleft}
\begin{tabular}{rccccrllrl}
\hline\hline\noalign{\smallskip}
Spectrograph&Epoch&No. of RVs&Dispersion&Wavelength range & Source\\
    No.   &(RJD=HJD-2400000)&prim/sec&(\AA mm$^{-1}$)&(\AA)    \\
\noalign{\smallskip}\hline\noalign{\smallskip}
 1&22177.8--22546.8& 21/22&29,49          &3900--5000&A\\
 1&22570.9--25841.0& 45/50&29,49,90       &3900--5000&B\\
 1&27284.8--28026.8& 35/35&14.8,21.5,29,49&          &D\\
 2&34527.8--36028.0& 27/27&36,40          &3800--4700&C\\
 3&36028.9--36360.9& 17/17& 10            &3800--6700&C\\
 1&36050.9--39116.6& 37/37& 14.8,29,49    &          &D\\
 4&39404.3--39991.4& 26/26& 37            &          &E\\
 5&44121.6--48042.8& 42/42& 0.9--1.4      &1500--    &F\\
 6&48105.4--49203.5& 42/42& 9.9           &4000--4680&G\\
 7&48142.8--48951.7& 29/29& $\sim10$      &4010--4970&H\\
\noalign{\smallskip}\hline\noalign{\smallskip}
 8&45897.8--45909.8& 14/08& 20            &3961--4510&I\\
 8&46305.0--48545.8& 29/28& 20            &3970--4510&I\\
 9&46694.8--47017.9& 09/09& 15            &3760--4195&J\\
10&51393.9--55220.6& 25/25& 10            &6150--6755&J\\
11&54027.3--54084.2& 02/02& 17.2          &6255--6767&J\\
\noalign{\smallskip}\hline\noalign{\smallskip}
\end{tabular}
\tablefoot{
{\it Col. 1:} \ \
1... Dominion Astrophysical Observatory 1.83 m reflector, 1-prism and 2-prism
     spg., IL, IM, IS, ISS and IIM configurations;
2... MtWilson 60-inch, $\gamma$ and X prism spgs.;
3... MtWilson 100-inch coud\'e grating spg.;
4... Crimea 1.22 m reflector;
5... International Ultraviolet Explorer, SPW high-dispersion spectra;
6... Calar Alto 2.2 m reflector, coud\'e spg., TEK CCD;
7... Kitt Peak Coud\'e Feed spg.;
8... Dominion Astrohysical Observatory 1.22 m reflector, coud\'e grating spg.,
     Reticon 1872RF;
9... Dominion Astrophysical Observatory 1.83 m reflector, Cassegrain grating
     spg., Reticon 1872RF;
10... Dominion Astrohysical Observatory 1.22 m reflector, coud\'e grating spg.,
      CCD SiTE-4 detector;
11... Ond\v{r}ejov Observatory 2.0 m reflector, coud\'e spg., CCD SiTE-5 \\
{\it Abbreviations used in Col. 6:} \ \
A... \citet{plaskett20}, remeasured by \citet{redman};
B... \citet{redman};
C... \citet{struve59}, partly remeasured by \citet{huffer};
D... this paper; unpublished RV measurements by Pearce and Petrie;
E... \citet{vitr};
F... \citet{stick09};
G... \citet{simon94};
H... \citet{burk97};
I... \citet{hill95} and this paper;
J... this paper
}
\end{flushleft}
\end{table*}

\begin{table*}
\caption[]{Photoelectric observations of \ye}
\label{jouphot}
\begin{flushleft}
\begin{tabular}{rcccccl}
\hline\hline\noalign{\smallskip}
Source & Station & Epoch & No. of obs.&No. of &HD$_{\rm comp.}$/& Passbands\\
       & No. &(RJD)& $U$/$B$/$V$ &nights &HD$_{\rm check}$ & used     \\
\noalign{\smallskip}\hline\noalign{\smallskip}
 1&74&33827.3--36071.4&347& 49&198820/197419&no filter ef. 4200 \AA\\
 2&76&36808.4--37589.4&155/29&7/2&198692/199007&filters 4550 \& 6000 \AA\\
 3&71&39712.3--40528.4&21/104/104&10&199007/ -- & $UBV$\\
 3&19&41119.4--41154.5& 48&  5&199007/ --   &$UBV$ \\
 3&42&41145.8--41190.8&  7&  3&199007/ --   &$UBV$ \\
 3&75&41159.4--41193.5&24/23&8&199007/ --   &$BV$ \\
 3&72&41169.7--42577.9&90/87/103&4/4/5&199007/ --   &$UBV$ \\
 4&70&44081.4--44083.7&127&  2&199007/ --   &$B$    \\
 7&94&47769.4--47788.5& 60&  6&199007/ --   &$V$\\
 5&61&47890.4--49040.7&141& 36& all-sky     &$V$ \\
 6&77&48040.1--48042.8& 33&  3& all-sky     &$V$ (IUE FES sensor)\\
 7&54&48178.3--48182.3&140&  5&199007/ --   &$V$\\
 8&01&51371.6--51377.4& 10&  4&204403/202349&$UBV$\\
 8&01&51379.4--51448.3&141& 13&202349/204403&$UBV$\\
 8&01&54011.3--54020.5& 93&  5&202349/199007, 198820&$UBV$\\
 8&01&55113.3--55126.4& 65& 13&199007/198692, 198820&$UBV$\\
 8&16&51452.6--51457.8& 79&  6&204403/199007&$UBV$\\
 8&16&53628.6--53667.8& 54& 12&204403/199007&$UBV$\\
 8&16&54008.8--54070.6& 85& 10&204403/199007&$UBV$\\
 8&16&54270.8--54285.9& 79& 22&204403/199007&$UBV$\\
 8&78&51477.0--51539.0&111&  4&199007/204403&$UBV$\\
 8&02&54025.2--54405.3&258&  3&199007/--    &$BVR$\\
\noalign{\smallskip}\hline\noalign{\smallskip}
\end{tabular}

{\it Abbreviations used in Col. 1:} \ \
1... \citet{maku};
2... \citet{herczeg72};
3... \citet{ocon};
4... \citet{alvaro80};
5... \citet{esa97};
6... \citet{stick09};
7... \citet{moss};
8... this paper.\\
{\it Abbreviations used in Col. 2 (numbers are running
numbers of the observing stations from the Praha / Hvar data archives):}\\
01... Hvar 0.65 m reflector, photoelectric photometer;
02... Brno private observatory of P.~Svoboda, Sonnar 0.135 m photographic lens
      and a CCD SBIG 7 camera;
16... Four College 0.80 m APT;
19... Abastumani 0.48 m;
42... Dyer Observatory 0.61 m Seyfert reflector, 1P21 tube;
54... Crimean 0.60 m reflector, EMI 9789 tube;
61... Hipparcos $H_p$ magnitude transformed to Johnson $V$ after \citet{hpvb};
70... Mojon de Trigo 0.32 m Cassegrain, EMI 6256 A tube;
71... Vatican 0.60 m Cassegrain, 1P21 tube;
72... Dominion Astrophysical Observatory 0.30 m, EMI 6256SA tube;
74... Abastumani 0.33 m;
75... T\"ubingen 0.40 m reflector;
76... Hoher List 0.35 m reflector;
77... International Ultraviolet Explorer: the fine error sensor
78... Sejong 0.40 m reflector;
94... Kazan station of the SAO 0.48 m reflector
\end{flushleft}
\end{table*}

There are several motivations for this study.
\begin{enumerate}
\item \cite{hill95} write in their section on photometry:
``We do not consider this photometric solution to be
definitive and hope that \yc will be observed again over a 4 year
span in a fully calibrated system." We circumvent this requirement
by combining our efforts from three stations, suitably separated in local
time. In 1999, we obtained calibrated $UBV$ photometry of the binary
from North America, Europe, and Korea. Additionally, we also have
new Reticon and CCD spectra of \yc at our disposal.
\item For such an astrophysically important system, it was deemed useful
to explore the effect of different physical assumptions and different
methods of analyses on the result and to see what accuracy in the determination
of the basic physical properties can actually be achieved.
\item We profit from having at our disposal the reduction procedures that
allow separate or simultaneous RV and light-curve solutions,
with the rate of apsidal advance as one of the elements of the solution.
This gives us the chance to use all available, as well as our new observations
in the determination of all basic physical elements of the binary with
an unprecedented accuracy and to define a more or less definitive orbit.
\item Thanks to the disentangling technique, we are also able
to test for the presence of rapid line-profile changes, which could be
related to forced non-radial oscillations.
\end{enumerate}
Here, we report our results.

\section{Observations and reductions}
\subsection{Spectroscopy}
Spectroscopic observations at our disposal consist primarily of
the following three series of electronic spectrograms obtained
at Dominion Astrophysical Observatory (DAO):
\begin{itemize}
\item 43 Reticon blue spectra already used by \cite{hill95};
\item 9 Reticon blue spectra obtained with the 1.83 m reflector and
      21121 spectrograph configurations by DH;
\item 19 red CCD spectra secured by SY with the 1.2 m reflector.
\end{itemize}
For further details on the DAO 21121, 21181 and 9681 spectrographs,
we refer the reader to \citet {rich}.

Additionally, we also used two CCD spectra secured by MW at the
coud\'e focus of the Ond\v{r}ejov 2 m reflector.

In all cases, calibration arc frames were obtained before and after
each stellar frame. During each night, series of ten flat--field and
ten bias exposures were obtained at the beginning, middle, and end of
the night.  These were later averaged for the processing of the stellar
data frames.  For the 1.83 m data, exposure times ranged from 15 to 30
minutes, with the signal-to-noise ratio ($S/N$) between 70 and 150,
while for the $1.2\,$ m data
exposure times of 20 minutes were used, giving $S/N$ between 32 and 180.
The data obtained by DH were re-reduced by SY with IRAF, from
initial extraction and flat--fielding up to wavelength calibration.
The Ond\v{r}ejov spectra were also reduced this way by M\v{S} in IRAF.
Continuum rectification of all spectra and cleaning from
cosmic ray hits (cosmics) was carried out
by PH using the program \spefo \citep[see][]{sef0,spefo}.
Following \citet{sef0}, we also measured a selection of good telluric lines
in all red spectra and used the difference between the calculated heliocentric
RV correction and the mean RV of these telluric lines to bring all red spectra
into one RV zero point.

To be able to derive the new, most accurate value of the rate of apsidal
advance, we also compiled all available RVs from the astronomical
literature, which can be found in Table~\ref{jourv}.
Here, and in several other tables we give the time instants
in an abbreviated form: RJD = HJD - 2400000.0.
We could not use three 1913-1914 Mt\,Wilson RVs published by \cite{abt73}
since they obviously refer to unresolved lines of both stars.
A few comments are appropriate. For the older DAO data, we adopted the
re-measurement by \citet{redman}. For those McDonald RVs that were
measured more than once, we calculated and adopted mean RVs for each
multiple measurement. The only exception was the spectrum
taken on Jul 7, 1957 at 10:31 UT for which \citet{huffer} measured
very deviating values. We adopted the original RVs from \cite{struve59}
for this particular spectrogram. We also omitted the McDonald spectrum
obtained on May 1, 1957 (HJD 2436324.9349), for which \citet{struve59}
tabulate a very deviant value, the same for both components (as also seen in
their original phase diagram). For the new spectra at our disposal,
the RVs were derived via Gaussian fits to the line
profiles of \ion{He}{I}~4026~\AA\ (blue spectra) and \ion{He}{I}~6678~\AA\
(red spectra).
Finally, for the electronic spectra at our disposal, we also derived
new orbital solutions using the \korel disentangling technique by
\citet{korel1,korel2,korel3,korel4} for several stronger lines
(as discussed in detail below).

\subsection{Photometry}
\begin{figure}
 \resizebox{\hsize}{!}{\includegraphics[angle=180]{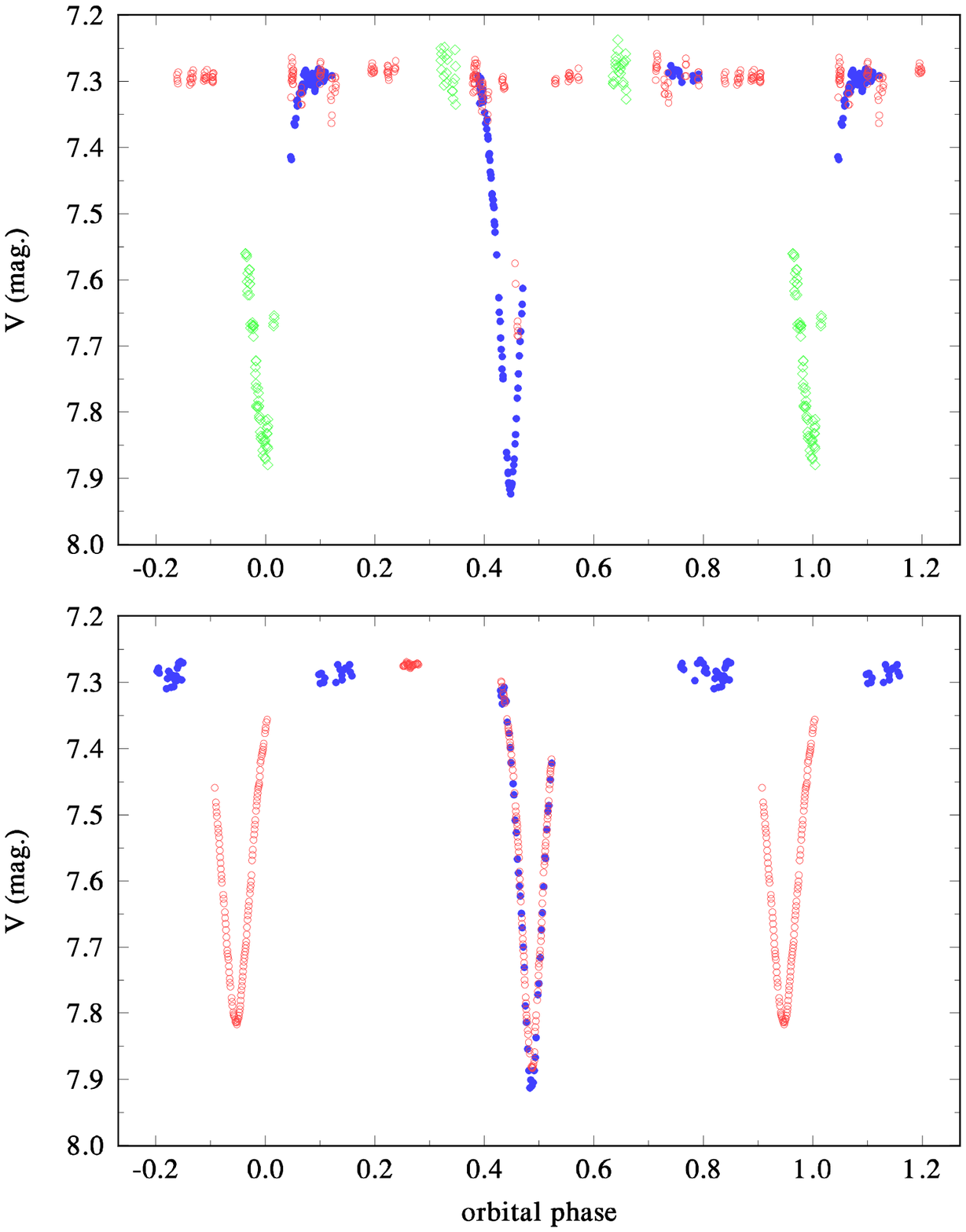}}
\caption{The $V-$band light curves: Upper panel shows the 1999 campaign
observations. Filled circles / blue in the electronic version denote the Hvar,
open circles / red the Villanova APT, and diamonds / green the Sejong
individual observations. The bottom panel shows the Hvar
(filled circles / blue) and Brno (open circles / red) 2006 and 2007
observations. The data are plotted vs. phase of the {\sl sidereal} orbital
period, with phase zero corresponding to one instant of the primary
mid-eclipse: HJD~2446308.3966.}
\label{lc}
\end{figure}

Several photoelectric light curves of \yc have been published.
The star has also been observed by Hipparcos and
new $UBV$ observations were secured during the 1999 observing campaign
in Hvar, Korea, and with the Four College APT in Arizona. Hvar observations
continued in 2006-2007 and 2009. In 2006, two eclipses in $BVR$
were observed by PS.
Basic information on all available data sets can be found in
Table~\ref{jouphot} and illustrative light curves based on our new
observations are shown in Fig.~\ref{lc}.

It was very unfortunate that different observers used different comparison
stars, even the red ones. However, most of the data sets were obtained
relative to HD~199007.

An important part of our new observations was, therefore, to observe all
these comparisons and to derive their good all-sky \ubv\ magnitudes.
Adding these to the magnitude differences variable minus comparison
allowed us to get many of the published data sets closer
to the standard $UBV$ system. More details on the data reduction and
homogenization can be found in the Appendix.

To derive new, improved values of the orbital and apsidal period,
we also used the historical light curve by \citet{dugan}, adopting $V$=9\m402
(based on the Hvar observations) for his comparison star
SAO~70600 = BD+33$^\circ$4062.

We provide all individual,
homogenised photometric observations together with their heliocentric
Julian dates in Table~4, which we publish in electronic form only.

\subsection{Times of minima}
We have collected all accurate times of primary and secondary mid-eclipses
available in the literature and complemented them
with the minima derived on the basis of our photometry.
 All photoelectric times of minima presented
in \citet{hol95} (their Table I) were included.
Over 500 reliable times of the primary and secondary minima
were used in our analysis. They now span an interval of about
125~years and are all listed in Table~5 (published in electronic form only).

\section{Towards a new accurate ephemeris}
To derive a new ephemeris including the rate of the apsidal advance,
we decided to derive and compare the results of three independent
determinations, based on (i) light-curve solutions, (ii) orbital
solutions, and (iii) analyses of the times of minima.

\setcounter{table}{5}
\begin{table}
\caption{Linear limb-darkening coefficients in the $UBV$ passbands
for stars with \tef near 30000~K and \lgg of 4.1 to 4.2
given by various authors}\label{limb}
\begin{flushleft}
\begin{tabular}{lccc}
\hline\hline\noalign{\smallskip}
Source       & $V$& $B$& $U$ \\
\noalign{\smallskip}\hline\noalign{\smallskip}
\citet{alnai78}     &0.27&0.32&0.32\\
\citet{wade}        &0.23&0.29&0.29\\
\citet{diaz}        &0.29&0.34&0.34\\
\citet{claret2000}  &0.33&0.37&0.38\\
\noalign{\smallskip}\hline\noalign{\smallskip}
\end{tabular}
\end{flushleft}
\end{table}

\begin{table*}
\caption[]{Exploratory \fotel and \phoebe light-curve solutions for all 4888
available yellow, blue, and ultraviolet observations with known times
of observations.}
\label{expl}
\begin{flushleft}
\begin{tabular}{lrrrrrr}
\hline\hline\noalign{\smallskip}
Element                       &     F1          &    F2          &    F3          &   F4       & P1\\
\noalign{\smallskip}\hline\noalign{\smallskip}
$P$ (d)                       &  2.99633161     &  2.99633161    &  2.99633162    &2.99633162  &2.99633179\\
$P_{\rm anomal.}$ (d)         &  2.99684586     &  2.99684586    &  2.99684587    &2.99684585  &2.99684593\\
                              &\p0.00000016     &\p0.00000016    &\p0.00000016    &\p0.00000016&0.00000070\\
$T_{\rm peri.}$ (RJD)         &  46308.66353    & 46308.66357    & 46308.66348    & 46308.66343&46308.66168\\
                              &    \p0.00038    &   \p0.00038    &   \p0.00038    &   \p0.00038&\p0.00018\\
$T_{\rm prim.ecl.}$ (RJD)     &  46308.39641    & 46308.39641    & 46308.39638    & 46308.39638&46308.39581\\
$e$                           &0.14514\p0.00020 &0.14512\p0.00020&0.14510\p0.00020&0.14511\p0.00020&0.14522\p0.00029\\
$\omega\,$ ($^\circ$)         &132.474\p0.050   &132.478\p0.050  &132.467\p0.051  &132.460\p0.051&132.280\p0.094\\
$\dot\omega$ ($^\circ\,{\rm d}^{-1}$)&  0.0206166&  0.0206168    &  0.0206169     &0.0206161&0.020626\\
                              &\p0.0000069      & \p0.0000069    &\p0.0000070     &\p0.0000070&\p0.000013\\
$r_1$                         & 0.2029\p0.0027  & 0.2033\p0.0030 & 0.2026\p0.0024 &0.2021\p0.0022&0.1954\\
$r_2$                         & 0.2040\p0.0017  & 0.2033\p0.0019 & 0.2043\p0.0015 &0.2049\p0.0014&0.2069\\
$i$ ($^\circ$)                & 86.587\p0.013   & 86.693\p0.012  & 86.619\p0.013  &86.571\p0.014&86.415\p0.029\\
\noalign{\smallskip}\hline
\end{tabular}
\tablefoot{
Solutions F1 to F4 are derived with \fotel for the linear limb-darkening
coefficients published by \citet{alnai78}, \citet{wade}, \citet{diaz}, and
\citet{claret2000}, respectively (see Table \ref{limb}). Solution P1 is
derived with the program \phoebe also using a linear law of the limb darkening
and adopting \citet{claret2000} values.
To save space, we only tabulate the basic elements of the solutions.}
\end{flushleft}
\end{table*}

To this end, we used two different computer programs, which are widely
used and allow the calculation of the light-curve and orbital solutions:
the program \fotele, developed by \citet{fotel1,fotel2}, and the program
\phoebe \citep{prsa05,prsa06} based on the WD\,2004 code \citep{wd71}.

We note that \fotel models the shape of the stars as triaxial
ellipsoids and uses linear limb-darkening coefficients while
\phoebe models the binary components as equipotential surfaces
of the Roche model and allows the use of non-linear limb darkening
laws. Both programs are, however, particularly well suited for our task
since they allow us to derive the rate of the periastron advance as one of
the elements.

We first carried out some exploratory solutions to map out the existing
uncertainties in the value of the orbital period and the rate of the
periastron advance as well as the sensitivity to different
limb-darkening coefficients found in the literature.
To this end, we first derived various trial solutions independently
for the photometric and RV data.

We used all photoelectric $\ubv$\ observations in combination with Dugan's
photometry. Altogether, this data set contains 4888 individual data points
spanning an interval from HJD~2420777.6 to 2454405.3 i.e. over 92~years.

\subsection{FOTEL exploratory light-curve solutions}
The values of the linear limb-darkening coefficients given in various
more recent studies are summarized in Table~\ref{limb}. We estimated
the values quoted from the paper by \citet{diaz} via interpolation
in their Table~1 after finding that their interpolation formula does not
perform particularly well near \tef = 30000~K.
To get some idea about how serious the uncertainties in the values of
the limb-darkening coefficients are, we derived four exploratory solutions
F1 to F4, using subsequently the linear limb-darkening coefficients
from the four sources given in Table~\ref{limb}.

\begin{table}
\caption[]{Exploratory \phoebe light-curve solutions for all 4888
available yellow, blue, and ultraviolet observations with known times
of observations, using square-root limb darkening coefficients.}
\label{sqroot}
\begin{flushleft}
\begin{tabular}{lrrrrrr}
\hline\hline\noalign{\smallskip}
Element                       &     P2          &    P3                 \\
\noalign{\smallskip}\hline\noalign{\smallskip}
$P$ (d)                       &  2.99633179     &  2.996331786          \\
$P_{\rm anomal.}$ (d)         &  2.99684596     &  2.996846065          \\
                              &\p0.00000070     &\p0.000000698          \\
$T_{\rm peri.}$ (RJD)         &  46308.66176    & 46308.66177           \\
                              &    \p0.00018    &   \p0.00017           \\
$T_{\rm prim.ecl.}$ (RJD)     &  46308.39582    & 46308.39576           \\
$e$                           &0.14521\p0.00029 &0.14508\p0.00029       \\
$\omega\,$ ($^\circ$)         &132.288\p0.094   &132.291\p0.094         \\
$\dot\omega$ ($^\circ\,{\rm d}^{-1}$)&  0.020614&  0.020618           \\
                              &\p0.000013       & \p0.000013           \\
$r_1$                         & 0.1959          & 0.1968                \\
$r_2$                         & 0.2064          & 0.2056                \\
$i$ ($^\circ$)                & 86.449\p0.028   & 86.693\p0.012         \\
\noalign{\smallskip}\hline
\end{tabular}
\tablefoot{
These solutions are derived with \phoebe for the square-root limb-darkening
law. For solution P2 the coefficients were adopted from \citet{claret2000},
while solution P3 is derived in such a way that in each iteration the program
interpolates in the new limb-darkening tables derived by Dr. A.~Pr\v{s}a
from the \citet{cas2004} model atmospheres.
To save space, we only tabulate the basic elements of the solutions.}
\end{flushleft}
\end{table}

In every case, we first derived a preliminary solution keeping the weights
of all data subsets equal to one. Then we weighted individual data sets by
the weights inversely proportional to the square of the rms error per
one observation from that solution and used them to obtain the final solution.
In practice, the weights ranged from 0.27 for Dugan's observations
to 20.6 for the best photoelectric observations.
All these exploratory solutions are summarized in Table~\ref{expl}.
Fortunately, it turned out that the difference between them
was much smaller than the respective rms errors of all elements derived.

\subsection{PHOEBE exploratory light-curve solutions}
For comparison, we also derived a joint exploratory solution,
based on all available light curves, using a recent {\sl devel} version
of the program \phoebe \citep{prsa05,prsa06} and also using linear
limb-darkening coefficients from \citet{claret2000}. This solution is
denoted P1 in Table~\ref{expl} and should correspond to solution F4
of that Table. Since \phoebe also allows the use of several non-linear
laws of limb darkening, we also compare (in Table~\ref{sqroot}) the \phoebe
solution for square-root limb-darkening coefficients adopted from
\citet{claret2000} (solution P2) and square-root law from limb darkening tables,
derived by Dr.~A.~Pr\v{s}a from the new \citet{cas2004} model atmospheres.
In this last case, the coefficients were interpolated after each iteration
for the current effective temperature (solution P3).

\begin{table}
\caption[]{Exploratory \fotel orbital solutions for all RVs found in the
astronomical literature and measured via Gaussian fits in our new spectra.
The numbering of the individual systemic velocities corresponds to that used
in Table~\ref{jourv}. We note, however, that since the RV zero point of the
red spectra was checked via measurements of a selection of telluric lines,
the two Ond\v{r}ejov spectra of spectrograph 11 were also treated as
belonging to spectrograph~9.}
\label{rvsol}
\begin{flushleft}
\begin{tabular}{lrrrrrr}
\hline\hline\noalign{\smallskip}
Element              & unweighted  & weighted  \\
\noalign{\smallskip}\hline\noalign{\smallskip}
$P_{\rm sider.}$ (d)             &    2.9963331   &    2.9963329 \\
$P_{\rm anomal.}$ (d)            &    2.9968425   &    2.9968458 \\
                                 &$\pm0.0000052$  &$\pm0.0000038$\\
$T_{\rm peri.}$ (RJD)            & 46308.620      &46308.637\\
                                 &   \p0.016      &  \p0.010\\
$T_{\rm prim.ecl.}$ (RJD)        & 46308.380      &46308.386\\
$e$                              &0.1335\p0.0050  &0.1404\p0.0032\\
$\omega\,$ ($^\circ$)            &127.4\p2.1      &129.6\p1.3\\
$\dot\omega\,$ ($^\circ$d$^{-1}$)&0.02042\p0.00022&0.02056\p0.00016\\
$K_1$                            &242.6\p1.6      &242.77\p0.99\\
$K_1/K_2$                        &1.0135\p0.0093  &1.0118\p0.0055\\
$\gamma_1$ (\ks)                 &$-$53.0\p  1.8  &$-$53.0\p 1.8 \\
$\gamma_2$ (\ks)                 &$-$62.6\p  4.0  &$-$62.6\p 4.0 \\
$\gamma_3$ (\ks)                 &$-$63.4\p  5.9  &$-$63.4\p 5.9 \\
$\gamma_4$ (\ks)                 &$-$58.9\p  4.2  &$-$58.9\p 4.3 \\
$\gamma_5$ (\ks)                 &$-$68.2\p  2.1  &$-$68.2\p 2.1 \\
$\gamma_6$ (\ks)                 &$-$63.2\p  1.2  &$-$63.1\p 1.2 \\
$\gamma_7$ (\ks)                 &$-$63.6\p  2.1  &$-$63.7\p 2.1 \\
$\gamma_8$ (\ks)                 &$-$52.6\p  2.0  &$-$52.7\p 2.0 \\
$\gamma_9$ (\ks)                 &$-$71.0\p  1.5  &$-$70.9\p 1.2 \\
$\gamma_{10}$ (\ks)              &$-$61.05\p 0.89 &$-$61.03\p0.85\\
rms (\ks)                        &23.8            &14.5 \\
\noalign{\smallskip}\hline
\end{tabular}
\end{flushleft}
\end{table}

\subsection{FOTEL exploratory orbital solutions}
We first derived the orbital elements from all available RVs assigning
weight one to all of them. This initial solution is denoted ``unweighted"
in Table~\ref{rvsol}. Then we assigned weights inversely proportional
to the rms error per one observation from this solution to all
ten subsets from different spectrographs and repeated the solution.
This improved solution is denoted ``weighted" in Table~\ref{rvsol}.
A few things are worth noting. It turned out that the RV measurements based
on the Gaussian fits are by far the most accurate of all classical
RV measurements. We also note a rather large scatter in the value of
the systemic velocity from different instruments. We do not think this
is indicative of a real change. It is more likely that this arises
from different investigators using different lines (and
perhaps different laboratory wavelengths). For O-type stars, the line blending
due to line broadening can be quite severe. For instance, the \ion{He}{I}~4026~\AA\ line is
partly blended with a \ion{He}{II} line, which probably explains the
most negative systemic velocity for spectrograph (spg.)~9. On the other hand,
the true systemic velocity is probably close to that of spg. No.~10;
the RVs are based on the Gaussian fit to the \ion{He}{I}~6678~\AA\ singlet
line.

\begin{table}
\caption[]{Another exploratory \fotel orbital solution
based on the most accurate \ion{He}{I}~6678~\AA\ RVs
measured via Gaussian fits in our new electronic spectra.}
\label{rvsolgau}
\begin{flushleft}
\begin{tabular}{lrrrrrr}
\hline\hline\noalign{\smallskip}
Element              & Value \\
\noalign{\smallskip}\hline\noalign{\smallskip}
$P_{\rm sider.}$ (d)             & 2.99633179 fixed \\
$P_{\rm anomal.}$ (d)            & 2.99684607 fixed \\
$T_{\rm peri.}$ (RJD)            &46308.6612\\
                                 &  \p0.0027\\
$T_{\rm prim.ecl.}$ (RJD)        &46308.3952\\
$e$                              &0.14508 fixed\\
$\omega\,$ ($^\circ$)            &132.291 fixed\\
$K_1$                            &245.9\p1.5\\
$K_1/K_2$                        &1.0210\p0.0092\\
$\gamma_{10}$ (\ks)              &$-$61.13\p0.84\\
rms (\ks)                        &6.12 \\
\noalign{\smallskip}\hline
\end{tabular}
\end{flushleft}
\end{table}

\begin{figure}
 \resizebox{\hsize}{!}{\includegraphics{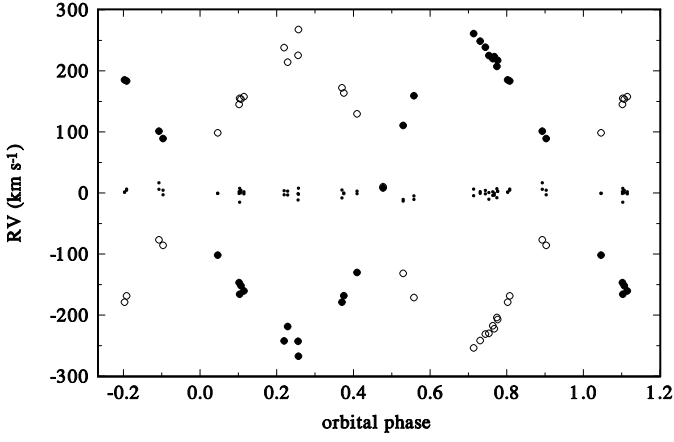}}
\caption{The RV curve based on the RVs from Gaussian fits to the
\ion{He}{I}~6678~\AA\ line from our new electronic spectra.
The data are plotted vs. phase of the {\sl sidereal} orbital period,
with phase zero corresponding to the primary mid-eclipse. The small dots
denote the residua from the orbital solution.}
\label{rvcgau}
\end{figure}

Since this last RV dataset is obviously the most accurate one, we also
present (in Table~\ref{rvsolgau}) another \fotel solution based on these
RVs only, keeping the orbital period, eccentricity, and the longitude
of periastron fixed from the solution P3 of Table~\ref{sqroot}.
A corresponding RV curve is in Fig.~\ref{rvcgau}.

\subsection{Apsidal motion from the analysis of the observed times of minima}
To compare different approaches, we analyzed all available times
of the observed primary and secondary minima.

\begin{table}
\caption{Elements derived from the analysis of the observed times
of the primary and secondary minima.}\label{apsidal}
\begin{flushleft}
\begin{tabular}{lccc}
\hline\hline\noalign{\smallskip}
Element      & Value         \\
\noalign{\smallskip}\hline\noalign{\smallskip}
$P_{\rm sider.}$ (days)    &2.99633210\p0.00000031    \\
$P_{\rm anomal.}$ (days)   &2.99684726\p0.00000031    \\
$T_{\rm prim.ecl.}$ (RJD)  &46308.39655\p0.00035 \\
$e$                        &0.1448\p0.0012  \\
$\omega$ (degrees)         &132.54\p0.18 \\
$\dot\omega$ (deg. per day)&0.020648\p0.00009\\
$U$  (years)               &47.73\p0.21  \\
\noalign{\smallskip}\hline\noalign{\smallskip}
\end{tabular}
\end{flushleft}
\end{table}

\begin{figure}
 \resizebox{\hsize}{!}{\includegraphics{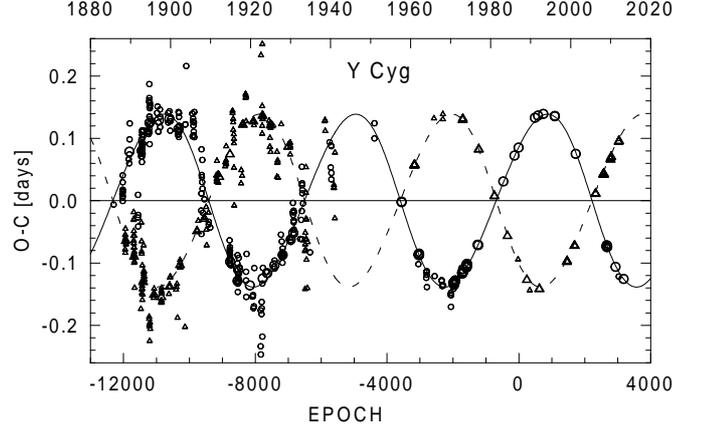}}
\caption{The \oc\ diagram for the times of minimum of Y~Cyg.
    The continuous and dashed curves represent predictions for the
    primary and secondary eclipses. The individual primary and
    secondary minima are shown as circles and triangles.
    Larger symbols correspond to the photoelectric or CCD measurements
    which were given higher weights in the calculations.}
\label{aps}
\end{figure}

To derive improved estimates of the periastron passage for
a chosen reference epoch $T_{\rm periastr.}$,
sidereal period $P_{\rm sider.}$, eccentricity $e$, longitude of
periastron at the reference epoch $\omega_0$, and the rate of apsidal
advance $\dot\omega$ (expressed in degrees per 1~day), we employed
the method described by \citet{gim83} and revised by \citet{gim95}.
This is a weighted least-squares iterative procedure, including terms
in the eccentricity up to the fifth power.
The periastron position $\omega$ is defined by the linear equation
\begin{equation}
\omega = \omega_0 + \dot{\omega}\ E,\label{ome0}
\end{equation}
\noindent where $E$ is the epoch of each recorded minimum.
The relation between the sidereal period $P_{\rm sider.}$ and
the anomalistic period $P_{\rm anom.}$ is given by
\begin{equation}
{1\over{P_{\rm anomal.}}} = {1\over{P_{\rm sider.}}} - {\dot\omega\over{360^\circ}},\label{pspa}
\end{equation}
\noindent and the period of apsidal motion $U$ in days by
\begin{equation}
U = {360^\circ\over{\dot\omega}}.\label{paps}
\end{equation}
\noindent
The orbital inclination $i = 86\fdg6$ was adopted on the basis
of our trial light-curve solutions. The resulting elements
with their errors are summarized in Table~\ref{apsidal} and the
corresponding \oc\ diagram is shown in Figure~\ref{aps}.

More details about the apsidal-motion solution and the comparison of
the present result with the previous studies of apsidal motion by
\citet{gim87} and \citet{hol95} can be found in a recent
conference paper by \citet{wolf2013}.

\subsection{Adopted ephemeris}
Comparing the results obtained in three different ways, which are summarized
in Tables~\ref{expl}, \ref{sqroot}, \ref{rvsol}, and \ref{apsidal},
we conclude that \phoebe solution P3 in Table~\ref{sqroot},
based on several thousands of individual photometric observations spanning
33628 days and the most detailed physical assumptions, provides
the best current ephemeris for the system. We adopt the period and the rate
of the apsidal advance from this solution and we use them consistently
in all the following analyses. We note that it implies the period of
apsidal motion $U=(47.805\pm0.030)$~yrs.

\subsection{Spectra disentangling for the electronic spectra}
As already mentioned, we used the program \korel developed by
\citet{korel1,korel2,korel3,korel4} for the spectra disentangling.
Preparation of data for the program \korel deserves a few comments.
The electronic spectra at our disposal have different spectral resolutions.
The DAO spectra with the dispersion of 20~\AAm\ (spg.~8 in
Table~\ref{jourv}) are recorded with a wavelength step of 0.3~\AA\
which translates to a RV resolution ranging from 22.7~\kms at the blue
end to 19.9~\kms at the red red. The 15~\AAm\ DAO spectra (spg.~9 of
Table~\ref{jourv}) have a 0.235~\AA\ separation between two consecutive
pixels which translates to 18.7 -- 16.8~\ks. The corresponding values
for the red DAO and Ond\v{r}ejov spectra are 0.150~\AA, i.e.
7.3 -- 6.5~\kms and 0.256~\AA\ and 12.3 -- 11.4~\ks.
We note that \korel and the Fast Fourier Transform (FFT) used in that program
require that the input spectra are rebinned into equidistant steps in RV in
such a way that the number of data points in each spectrum is an integer
power of 2. We also note that the two subsets of the blue DAO Reticon
spectra cover two overlapping spectral regions. For this reason,
we investigated the following three spectral regions with \korele,
using the RV steps to satisfy the above conditions:

\smallskip
4003 -- 4180 \AA, RV step 14.2 \ks,

4175 -- 4495 \AA, RV step 11.5 \ks, and

6340 -- 6740 \AA, RV step 4.75 \ks.

\smallskip
\noindent We omitted the blue DAO spectra r3049 to r4136 (RJDs 46304.97
to 46330.78) that were used by \citet{hill95}, because there was a~technical
problem with the Reticon detector at that time and the spectra are very poor
and strongly underexposed. We also omitted another very poor
spectrum r7712 (RJD~46611.79).
The rebinning was carried out with the help of the program {\tt HEC35D}
written by PH\footnote{The program {\tt HEC35D} with User's Manual
is available to interested users at
{\sl ftp://astro.troja.mff.cuni.cz/hec/HEC35}}
which derives consecutive discrete wavelengths via
\begin{equation}
\lambda_{\rm n}=\lambda_1\left(1+{\dl RV\over{c}}\right)^{\rm n-1},\label{delrvc}
\end{equation}
\noindent where $\lambda_1$ is the chosen initial wavelength,
$\dl RV$ is the constant step in RV between consecutive wavelengths,
and $\lambda_{\rm n}$ is the wavelength of the $n$-th rebinned pixel.
Relative fluxes for these new wavelength points are derived using
the program INTEP \citep{hill82}, which is a modification of
the Hermite interpolation formula. It is possible to choose
the initial and last wavelength and the program smoothly fills in
the rebinned spectra with continuum values of 1.0 at both edges.

To take the variable quality of individual spectra into account, we measured
their $S/N$ ratios in the line-free region 4170--4179~\AA\ for the blue
spectra, and 6635--6652~\AA\ for the red spectra, and assigned
each spectrum a weight according to formula

\begin{equation}
w={(S/N)^2\over{(S/N_{\rm mean})^2}},\label{wk}
\end{equation}

\noindent where $S/N_{\rm mean}$ denotes the mean $S/N$ ratio of all spectra.
Specifically, the $S/N$ ratio ranged between 42 and 355 for the blue, and
between 66 and 262 for the red spectra. We note that \korel uses the observed
spectra and derives both, the orbital elements
and the mean individual line profiles of the two binary components.

\begin{table}
\caption[]{KOREL disentangling orbital solutions for three spectral
regions.}\label{korel}
\begin{flushleft}
\begin{tabular}{lllllllllllllll}
\hline\hline\noalign{\smallskip}
Element&4003--4180 \AA     &4175--4495 \AA     &6340--6740 \AA \\
\noalign{\smallskip}\hline\noalign{\smallskip}
$T_{\rm periastr.}$&46308.6618&46308.6562&46308.6635\\
$e$                &0.14448&0.14929&0.14640\\
$\omega\,(^\circ)$&132.35&132.42&132.38\\
$K_1$(\ks) &242.72 &244.63 &246.77 \\
$K_2$(\ks) &247.68 &241.42 &244.69 \\
$K_1/K_2$ &0.9800 &1.0133&1.0085\\
\noalign{\smallskip}\hline
\end{tabular}
\end{flushleft}
\tablefoot{The anomalistic period, the rate of periastron advance, and
eccentricity were kept fixed at values of 2\D996845, 0.020645$\,$degrees
per day, and 0.1451. All epochs are in RJD.}
\end{table}

\begin{figure}
\begin{minipage}[t]{0.49\linewidth}
\centering
 \includegraphics[width=\textwidth]{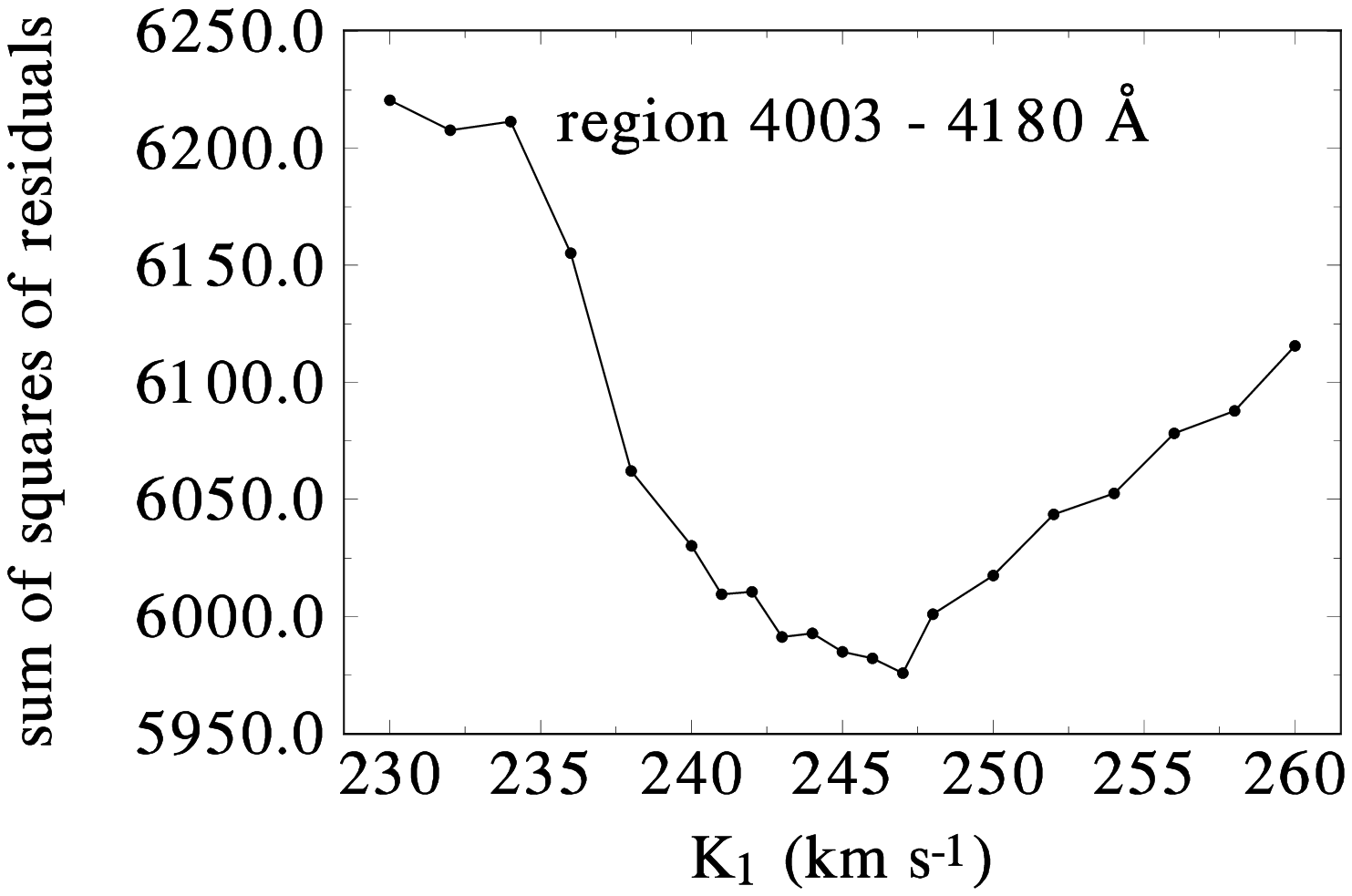}
 \includegraphics[width=\textwidth]{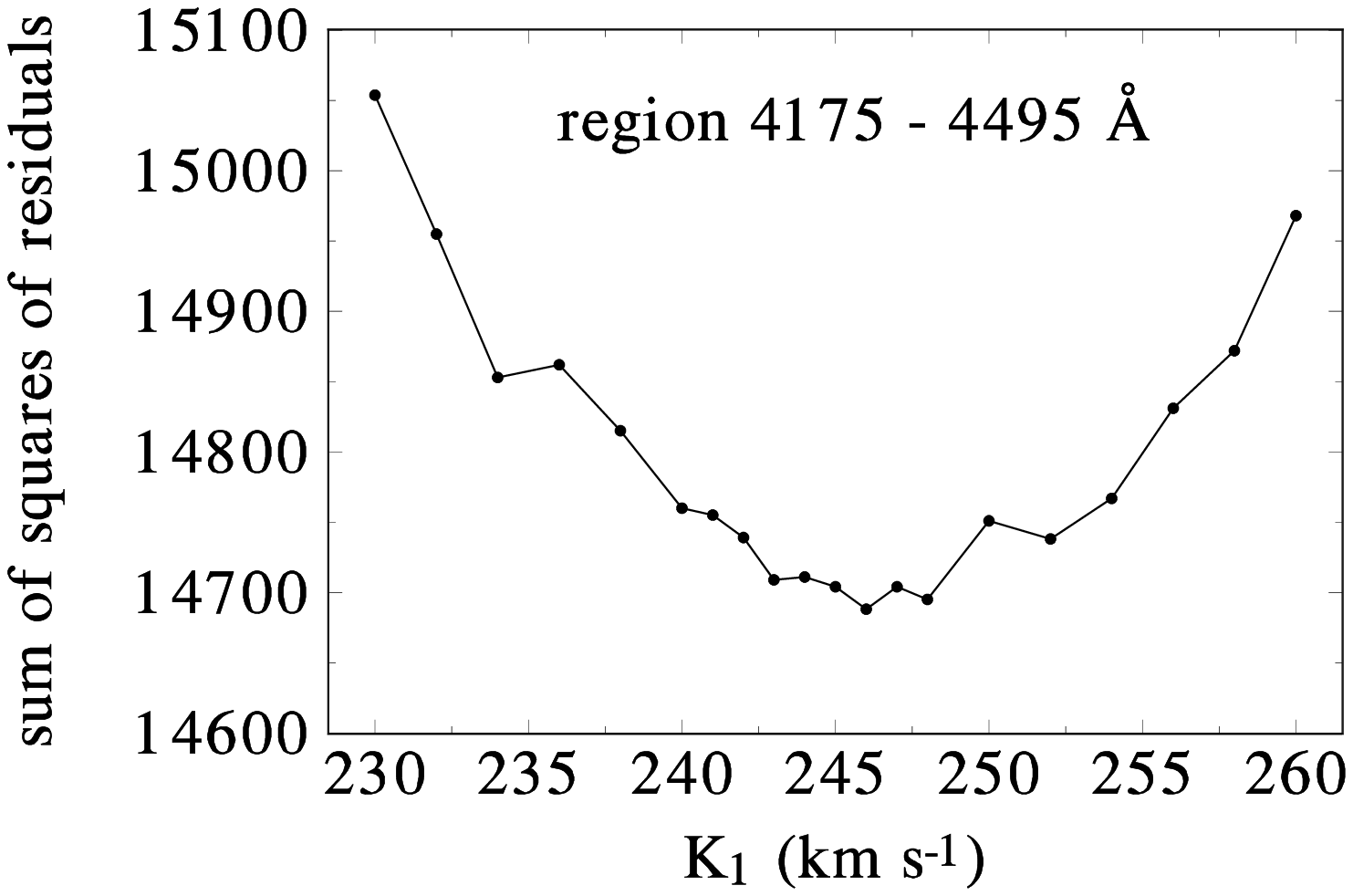}
 \includegraphics[width=\textwidth]{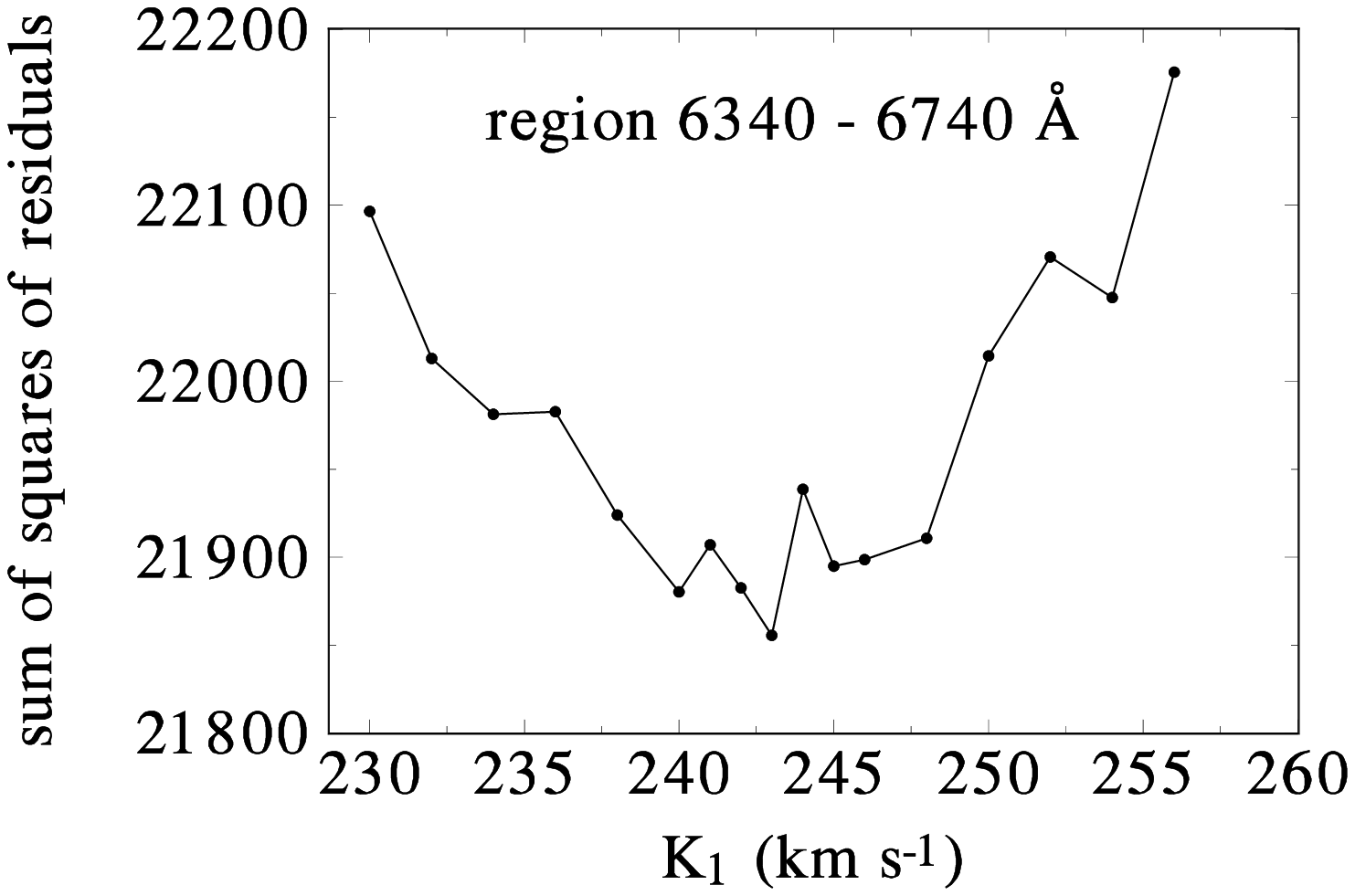}
\caption{Search for the best value of the semiamplitude of star~1
with \korel in three spectral regions, keeping the mass ratio fixed
at the value of 1.012.}
\label{k1map}
\end{minipage}
\begin{minipage}[t]{0.49\linewidth}
\centering
 \includegraphics[width=\textwidth]{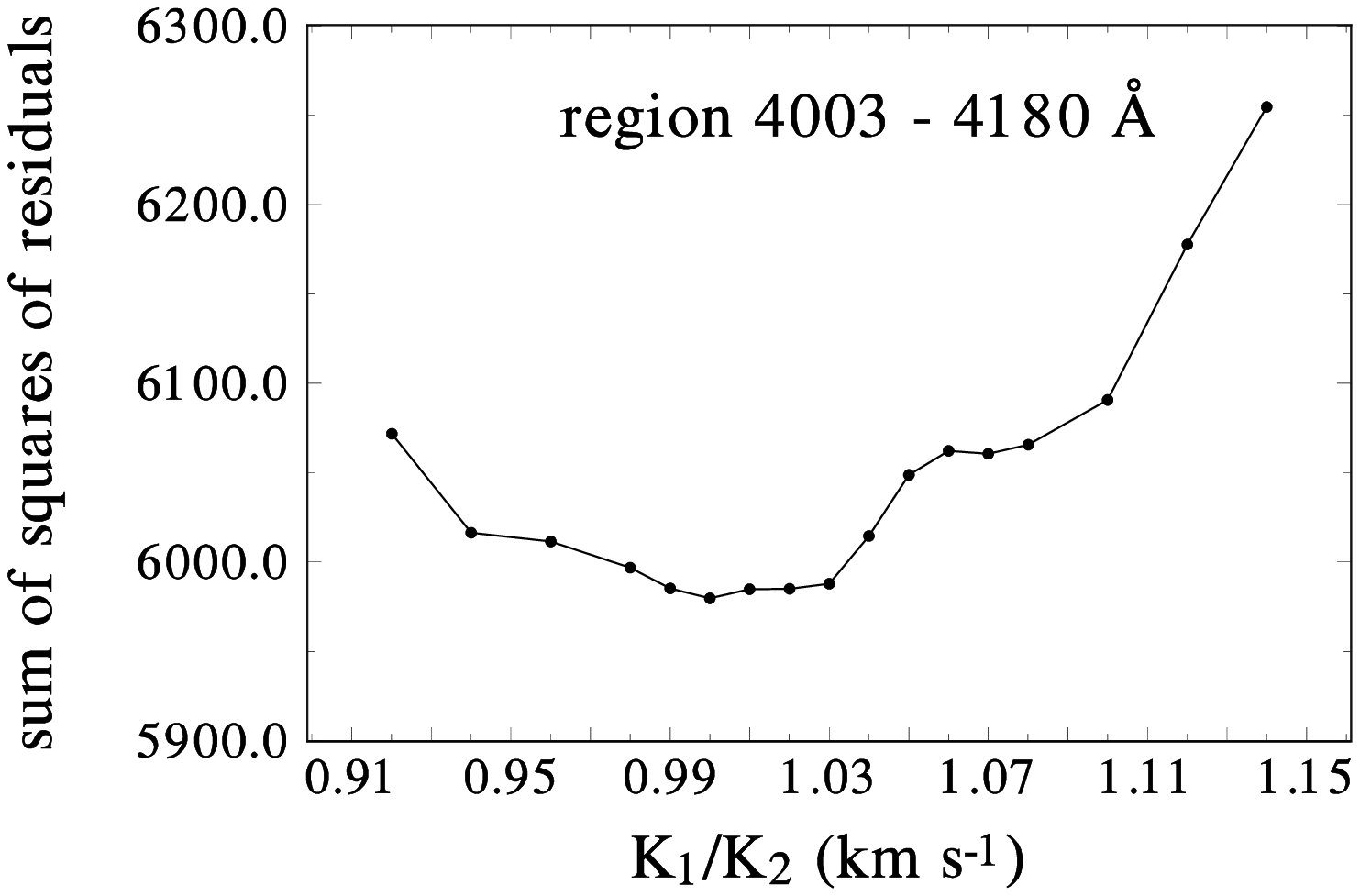}
 \includegraphics[width=\textwidth]{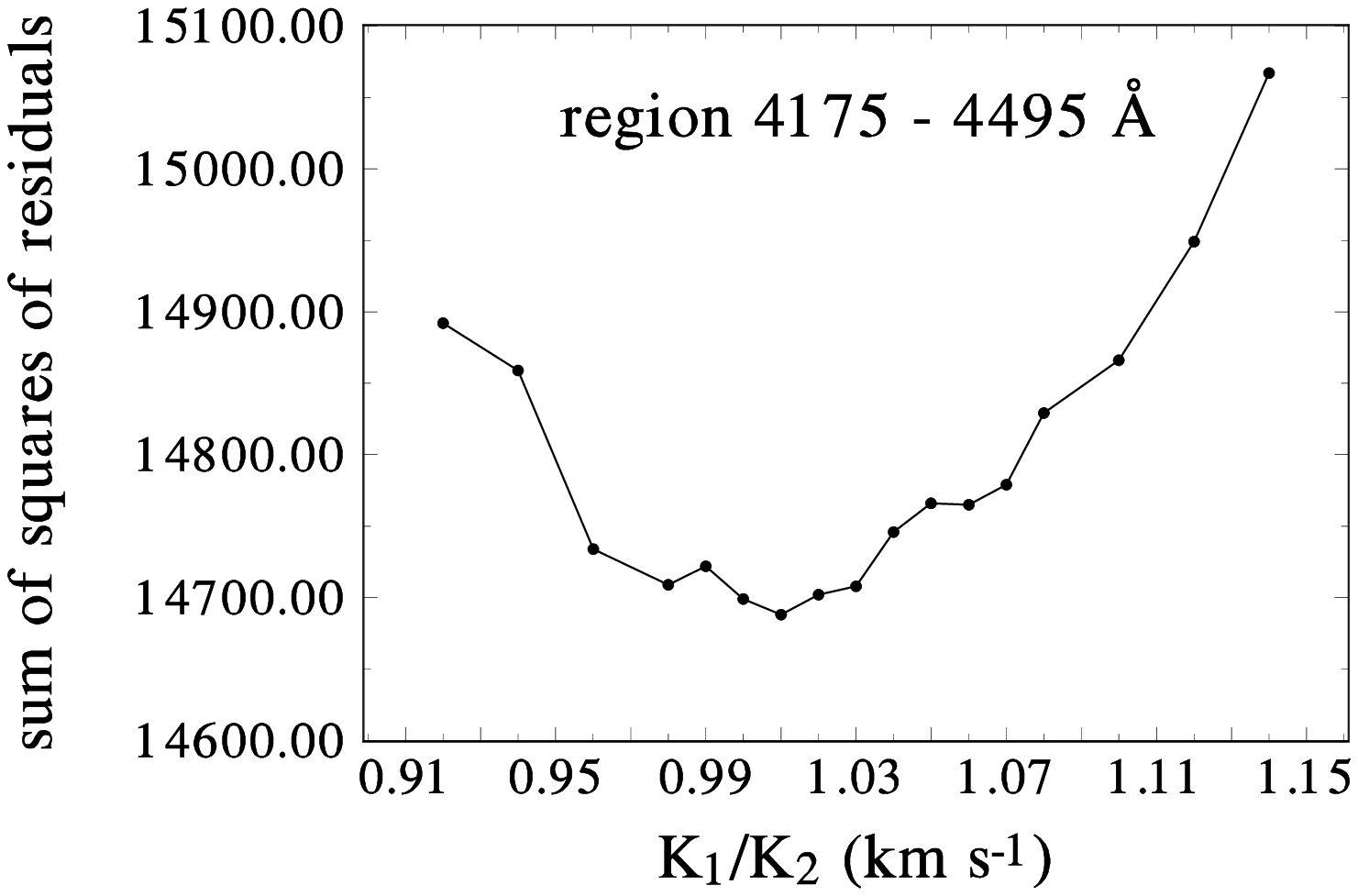}
 \includegraphics[width=\textwidth]{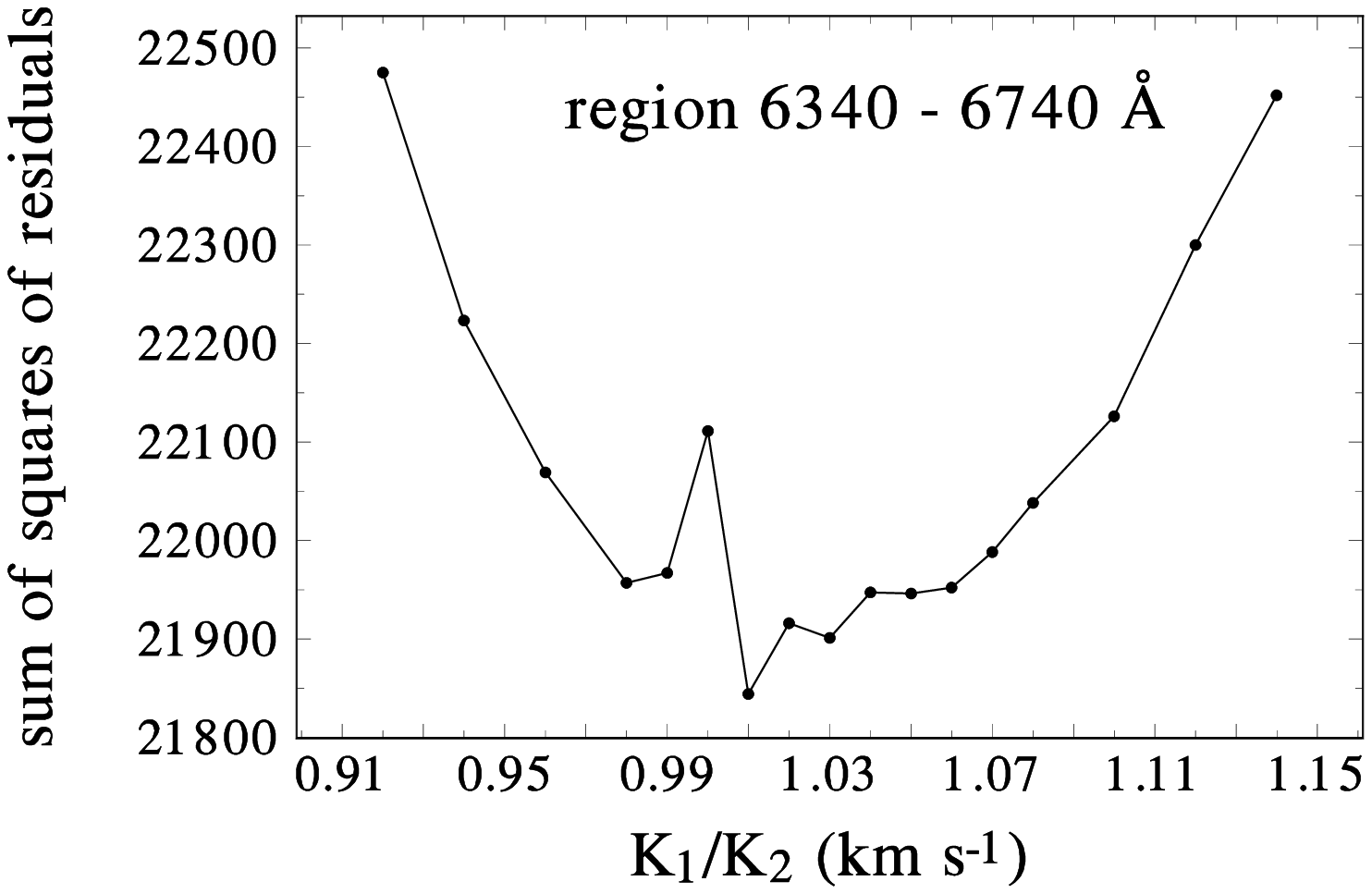}
\caption{Search for the best value of the mass ratio $K_1/K_2$
with \korel in three spectral regions, keeping the semi-amplitude
of the star~1 equal to 243, 245, and 246~\ks, respectively (resulting
from the minima of the $K_1$ maps).}
\label{qmap}
\end{minipage}
\end{figure}

We first investigated the variation of the sum of squares as a function
of the semiaplitude of star~1 and the mass ratio, considering our
experience that the sum of squares has a number of local minima.
These maps for the semiamplitude $K_1$ and the mass ratio $q=K_1/K_2$
for all three considered spectral regions are shown in Figs.~\ref{k1map}
and \ref{qmap}, respectively. They show that the most probable value of
$K_1$ is around 245~\ks; for $q$ this is close to 1.01.
The final \korel solutions for the three available wavelength intervals
were then derived via free convergency starting with the values
close to those corresponding to the lowest sum of squares found
from our two-parameter maps. We kicked the parameters somewhat from
their optimal values and ran a number of solutions to find the one
with the lowest sum of residuals in each spectral region.
In the initial attempts, we verified that the values of $e$ and $\omega$
from the \korel solutions agree reasonably with those from the photometric
solutions. To decrease the number of free paramaters, we also fixed
the value of the orbital eccentricity from the photometric solutions
at 0.1451 and ran another series of \korel solutions to find those with
the lowest sum of squares of residuals. These were adopted as final and
are summarized in Table~\ref{korel}.

\begin{figure}
 \resizebox{\hsize}{!}{\includegraphics{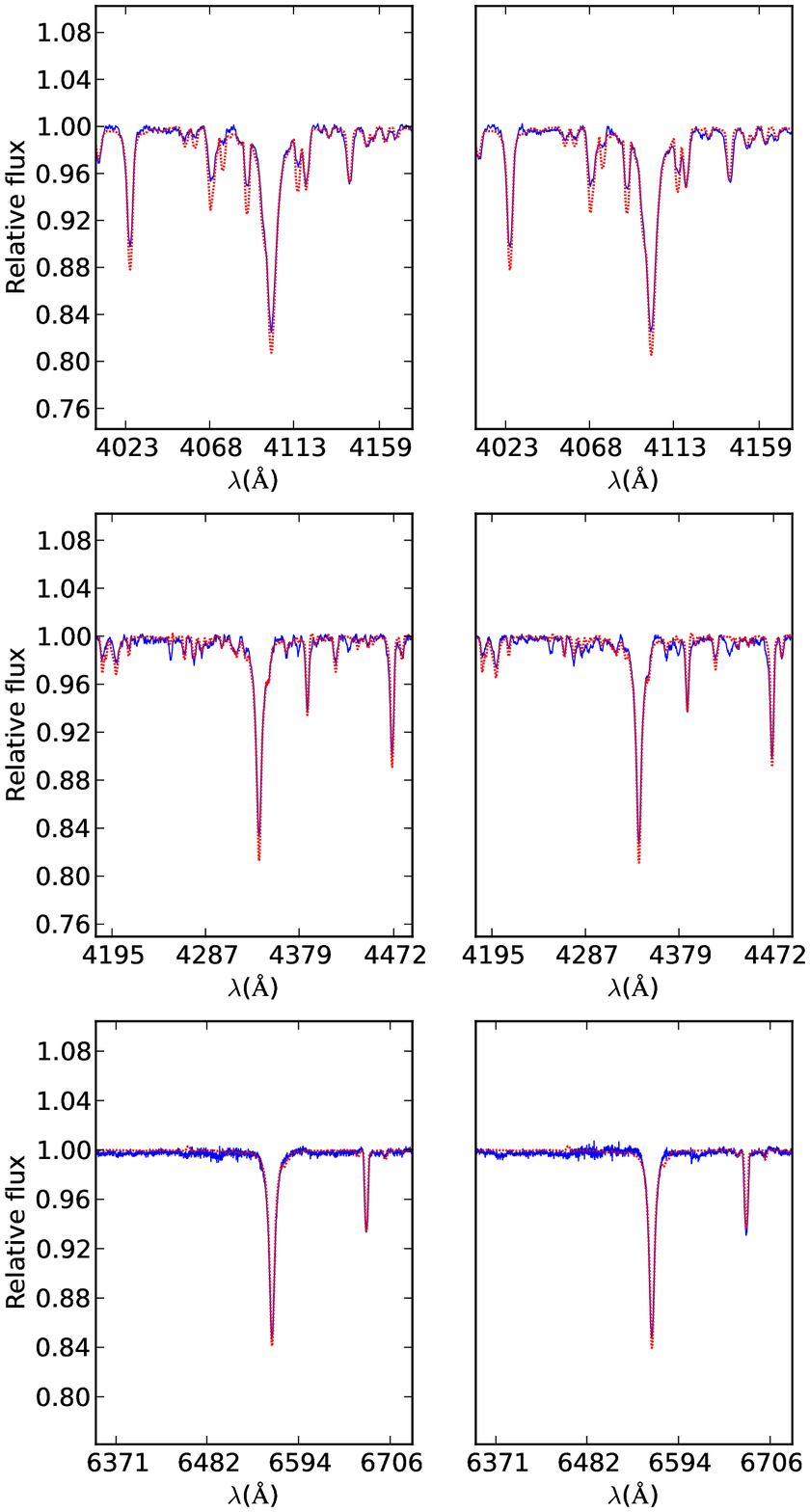}}
\caption{The comparison of the KOREL disentangled spectra (solid / blue line)
with synthetic spectra (dotted / red thin line). The left panels show
the spectrum of the primary compared to the synthetic spectrum
with \tef = 33200~K, \lgg = 4.161 [cgs], vsin =132 \ks, and a relative
luminosity 0.496. The right panels show the
spectrum of the secondary compared to the model spectrum
with \tef = 33521~K, \lgg = 4.17 [cgs]
broadened to \vsin = 132~\ks, and a relative luminosity 0.504.}
\label{prof}
\end{figure}

\section{Final elements}

\begin{table*}
\begin{center}
\vspace{0.1cm} \caption[ ]{The final solution and absolute elements.
$\Omega$ is the value of the Roche-model potential used
in the WD program, and $L_i$ are the relative luminosities
of the components in individual photometric passbands. They are normalised in
such a way that $L_1+L_2=1$.The system magnitudes at maximum light
$V_{1+2}$, $B_{1+2}$, and $U_{1+2}$ are based on the \phoebe model light
curves for the calibrated Hvar photometry.}\label{final}
\begin{tabular}{ll|ccl|lcc}
\hline\hline\noalign{\smallskip}
&& \multicolumn{3}{c|}{Combined \korel and \phoebe solution}\\
Element &                & Primary &System & Secondary \\
\noalign{\smallskip}\hline\noalign{\smallskip}
$a$                       &$(R_{\sun})$ && 28.72 fixed   & \\
$q$                       &             && 1.001 fixed   & \\
$T_{\rm periastr.}$       &(RJD)        &&$46308.66407\pm0.00010$&\\
$T_{\rm min.I}$           &(RJD)        &&$46308.39658\pm0.00010$&\\
$e$                       &             &&0.14508 fixed&\\
$\omega$                  &$({}^\circ)$ &$132.514\pm0.052$&&312.514\\
$i$                       &$({}^\circ)$ &&$86.474\pm0.019$&   \\
$\Omega$                  &             &$6.186\pm0.014$ && $6.162\pm0.014$\\
$T_\mathrm{eff}$          &(K)          &$33200\pm200$ fixed&& $33521 \pm 40$\\
$M$                       &($M_{\sun}$) & $17.72 \pm 0.35$ && $17.73 \pm 0.30$\\
$R$                       &($R_{\sun}$) & $5.785 \pm 0.091$ && $5.816 \pm 0.063$\\
$M_\mathrm{bol}$          &(mag)        & $-6.65 \pm 0.04$ && $-6.70 \pm 0.04$ \\
$\lgg$                    &[cgs]        & $4.161 \pm 0.014$ && $4.157 \pm 0.010$ \\
$L_i$ & V band & $0.496\pm0.003$ & & 0.504  \\
$L_i$ & B band & $0.496\pm0.003$ & & 0.504  \\
$L_i$ & U band & $0.495\pm0.003$ & & 0.505  \\
$V_{1+2}$&(mag.)&                 &7.2845&   \\
$B_{1+2}$&(mag.)&                 &7.2296&   \\
$U_{1+2}$&(mag.)&                 &6.3137&   \\
$V_0$    &(mag.)&7.287 & & 7.265             \\
(\bv)$_0$&(mag.)&$-$0.291 & &$-$0.291        \\
(\ub)$_0$&(mag.)&$-$1.086 & &$-$1.091        \\
$M_{\rm V}$&(mag.)&$-$3.59 & &$-$3.62        \\
\noalign{\smallskip}\hline\noalign{\smallskip}
\end{tabular}
\end{center}
\end{table*}

To derive the final solution leading to the improved values of the basic
physical properties of the binary components and of the system, we had
to proceed in an iterative way. One of us (JN) has developed a program that
interpolates in the grid of synthetic spectra and by minimizing the difference
between the observed and model spectra, estimates the optimal values
of $T_{\rm eff},$ $\log~g,$ $v\sin~i$, and RV. The program was applied to disentangled
spectra of both binary components using the NLTE line-blanketed spectra
from the O-star grid published by \citet{lanz2003}. We then used the resulting
\tef\ for star~1 and kept it fixed in the light-curve solution in
the program \phoebee. This time, we used only the \ubv\ observations that we
transformed to the standard Johnson system i.e. the data from stations
1, 16, and 78 plus two good-quality sets of $BV$ observations
(stations 71 and 75) and the $R$ photometry from station 2 in
Table~\ref{jouphot}.

We note that \korel also allows the determination
of RVs of individual components. However, when these RVs are used in \fotele,
they lead to an orbital solution, which differs somewhat from that derived by
\korele. The reason is that the RVs are derived
from individual spectra (of varying $S/N$ ratio),
while the orbital elements in \korel are given by the whole ensemble
of the spectra, weighted by the square of their $S/N$ ratio (cf. eq.~\ref{wk}).

We therefore adopted the mean values from the three solutions
in Table~\ref{korel},

\smallskip
\centerline{$K_1=244.7\pm2.0$ \ks, $K_2=244.6\pm3.1$ \ks,}
\centerline{$q=1.001\pm0.015,$}

\smallskip
\noindent to derive the semi-major axis $a = (28.72\pm0.22)$~\rs. It was then,
together with $q$, fixed in \phoebe to reproduce the \korel orbital solution.

For the eccentricity of 0.14508 one can estimate that the synchronization
at periastron would require the angular rotational speeds of the binary
components to be 1.354  times faster than the orbital angular speed. We
fixed this value in \phoebee.
Using the radii of the components resulting from the \phoebe solution, we
estimate \vsin = 132~\kms for both binary components.

 When \phoebe converged, we used the resulting
\lgg and the relative luminosities $L_1$ and $L_2$ (expressed in units
of the total luminosity outside eclipses) for both components and kept
them fixed for another fit of disentangled spectra. A new value
of $T_{\rm eff1}$ was then used in \phoebee. We arrived at a final set
of elements after three such iterative cycles.
The final solution is presented in detail in Table~\ref{final} and the
disentangled and model spectra are compared in Fig.~\ref{prof}.

The dereddened colours of the binary components correspond well to their
spectral class O9.5 and the solution implies a distance modulus of 10.9,
consistently for both binary components. The corresponding parallax
of $0\farcs00066$ does not contradict the Hipparcos revised parallax
of $0\farcs00080\pm0\farcs00046$ \citep{leeuw2007a, leeuw2007b}.

\section{A negative search for rapid line-profile variations}
To check whether or not line profile variability (LPV) is present in either
component of Y Cyg, we first used \korel to generate difference spectra
for each component.  This was done once the solutions had been converged
properly. Essentially, \korel shifts a recovered component line profile
to the appropriate radial velocity, and then subtracts the line profile
from the observed spectrum.  If any LPV is present, it should be easily
detectable in these difference spectra.  The temporal variance spectrum TVS
\citep{alex96} was used to search for any signature of LPV in the
He~I~4388 line.  This line was chosen because there are no nearby strong
blends.  Upon investigating both TVS plots, we conclude that no
LPV, detectable at the level of accuracy of our data, is present.
If LPVs {\it were} present, one would see sharp peaks in the TVS rising
above the noise level at the location of the line, which is not the case.

\section{Conclusions}
The results of a very detailed study of a huge body of spectral and
photometric observations of an archetypal early-type eclipsing binary
with a fast apsidal motion can be summarized as follows:

\begin{figure}
 \resizebox{\hsize}{!}{\includegraphics{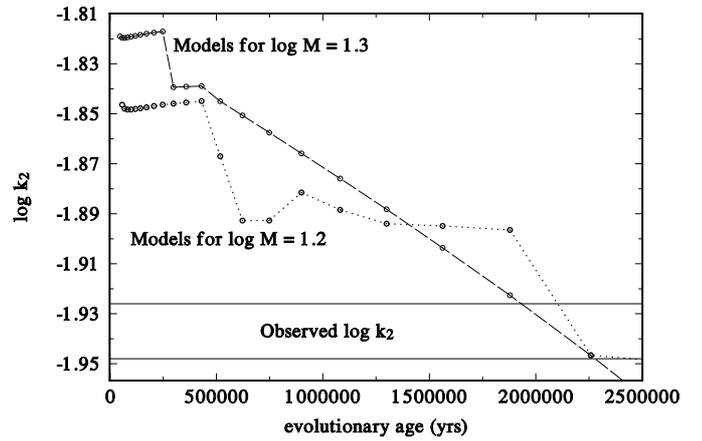}}
\caption{The comparison of the observed range of the logarithm of
the internal structure constant with its theoretical value based
on the evolutionary models of \citet{claret2004}, calculated for
log~M 1.2 and 1.3 (masses 15.85~\ms\ and 19.95~\ms, respectively).
The observed and theoretical values agree for an evolutionary age
slightly over 2\,000\,000 years.}
\label{logk}
\end{figure}

\begin{enumerate}
\item A comparison of separate analyses of radial-velocity and photometric
observations and of recorded times of minima covering about 130 years
shows that mutually consistent determinations of the sidereal and anomalistic
period and the period of apsidal-line rotation are obtained. The most accurate
results are based on the \phoebe analysis of all complete light curves.
\item The new values of masses, radii, effective temperatures,
and luminosities of the binary components are now quite robust and indicate
a close similarity of both bodies. However, it is a bit disappointing that these
new solutions do not represent a substantial improvement over the previously published
determinations. The problem arises partly from the fact that all electronic
spectra at our disposal have a moderate spectral resolution (from 13000
to 20000), and partly in inherent problems on the side of theory.
For instance, we were unable to achieve a perfect match between
the disentangled and synthetic spectra. In addition, the true uncertainties in
the estimated effective temperatures are surely much higher than the formal
errors in the final solution. We note that \citet{hill95} estimated effective
temperatures  $31000\pm2000$~K and $31600\pm2000$~K and \lgg = 4.00 [cgs] and
4.00 [cgs] for stars~1 and 2, respectively. \citet{simon94b}
carried out a detailed comparison of the observed and synthetic NLTE spectra
derived by \citet{kunze90, kunze94} and arrived at $T_{\rm eff1} = 34200\pm600$~K,
$T_{\rm eff2} = 34500\pm600$~K, log~$g_1 = 4.18$ [cgs],
and log~$g_2 = 4.16$ [cgs]. As pointed
out to us by Hubeny (2013, priv. comm.), the theoretical spectra computed by Kunze
do not include the metal line blanketing. This indicates that they probably
{\sl overestimate} the effective temperatures by as much as $\sim$2000~K.
Even when one compares the synthetic spectra from the O-star grid,
and the later published spectra from the B-star grid \citep{lanz2007} in the
overlapping temperature range, it turns out that the two sets of model
spectra differ from each other since the B-star values are based on somewhat
improved input physics. Clearly, without further progress in the model spectra,
one cannot do any better.
\item We also note that our new masses, radii, and effective
temperatures, together with the observed rate of the apsidal motion,
imply the relativistic contribution to the apsidal motion
$\dot\omega_{\rm rel.} = 0.0009645\,\,^\circ$d$^{-1}$, i.e. about 4.6~\% of
the total observed rate of the apsidal rotation. This implies the
observed value of the logarithm of the internal structure constant
(average of the two nearly identical stars)
$\log~k_2 = -1.937\pm0.011.$ In Fig.~\ref{logk} we compare the range
of this observed value (within the quoted error) with the evolutionary
models of \citet{claret2004}.  The observed and theoretical values agree
with each other for an evolutionary age slightly over $2\times10^6$~yrs.
\end{enumerate}

\begin{acknowledgements}
DH would like to thank the director and staff of the Dominion
Astrophysical Observatory for the generous allocation of
observing time and technical assistance.
We thank Dr. L.V. Mossakovskaya who kindly provided us with her unpublished
$V$ observations of two minima secured in 1989 and 1990. Dr.~\v{Z}.~Ivezi\'c
kindly helped us to obtain a copy of Dugan's paper with the visual photometry
of Y~Cyg.
The use of the following internet-based resources is gratefully acknowledged:
the SIMBAD database and the VizieR service operated at CDS, Strasbourg, France;
the NASA's Astrophysics Data System Bibliographic Services, and
the \oc\ gateway of the Czech Astronomical Society.~\footnote{\tt
http://var.astro.cz/ocgate/} In the initial stage, this study
was supported by the grant A3003805 of the Grant Agency of the Academy of
Sciences of the Czech Republic. Later, it was supported by the
grants GA~\v{C}R 205/03/0788, 205/06/0304, 205/06/0584,
and 209/10/0715 of the Czech Science Foundation and also by
the research projects AV0Z10030501 and MSM0021620860.
E.F.~Guinan and G.~McCook wish to acknowledge support
from the US National Science Foundation Grants NSF/RUI AST-0507536 and AST-1009903.
\end{acknowledgements}

\bibliographystyle{aa}
\bibliography{citace}

\Online
\begin{appendix}
\section{Details of the reduction and transformation
          of photometric data}
Comments on individual data sets:

{\it Station 19: Abastumani 0.48 m reflector} \\
These observations are on an instrumental $UBV$ system and suffer from
an unusually large scatter. Moreover, the last night (HJD~2441154) seems
to deviate from the general light curve quite a lot. In addition, the times of
the inflection inside the light minimum differ notably for different
passbands for this particular night. Since we now have at our disposal
numerous observations covering a similar period of time, we excluded
these definitely lower-quality data from our analyses.

{\it Station 42: Dyer} \\
A very limited set of instrumental $UBV$ observations from three nights only.

{\it Station 70: Mojon de Trigo} \\
The data for the night HJD~2444083 in Table~I of \citet{alvaro80} are not
presented in an increasing order which looks a bit confusing, but there
is no actual misprint in the table, as kindly communicated to us by
Dr.~Alvaro~Gim\'enez. The data are on an instrumental system.

{\it Station 71: Vatican} \\
According to \citet{ocon}, these data were transformed to the standard
Johnson system using observations of several $UBV$ standards. In addition,
the extinction was derived and appropriate corrections were applied.
We note that there is an obvious misprint in the first observation,
obtained on HJD~2439712.3492, at the $U$ colour: It should read
-0\m641 instead of the tabulated value of -0\m941.

{\it Station 72: University of Victoria 0.30 m} \\
These are data on an instrumental $UBV$ system. No extinction corrections
were applied.

{\it Station 74: Abastumani 0.33 m} \\
This is the largest homogeneous set of photoelectric observations of \ye,
covering the period from 1950 to 1957. The data were obtained without
any filter and the check star used, HD~197419, is listed under the name
V568~Cyg in the General Catalogue of Variable Stars. There is a misprint
in the last page of Table~1 of \citet{maku}: the first Julian date is
HJD~2435663 (continuation from the previous page), not 2435659 as given
in the Table. The data are of a good quality, with the exception of the
following deviating observations which we omitted from our analysis:
HJD~2434209.316, all three observations from 2434897, 2434949.298,
2435635.408, .423, 2435647.365, 2436071.339, and .342.

{\it Station 75: T\"ubingen 0.40 m} \\
A limited set of instrumental $BV$ observations.

{\it Station 76: Hoher List} \\
These early blue and yellow observations are on an instrumental system.
They were secured relative to a red comparison star HD~198692.

{\it Station 77: IUE Fine Error Sensor} \\
These observations, transformed to Johnson $V$ by \cite{stick09},
come from two tracking stations. They have larger scatter than
most of the photoelectric observations secured on the Earth, but their
time distribution ensures a relatively uniform, although not dense
coverage of the light curve. There is one deviating point at
HJD~2448040.442 that we omitted from the analyses.

\begin{table*}
\caption{Standard $UBV$ magnitudes of all comparison stars used by various
investigators in their differential photometries of \ye. The values
we derived from the calibrated and carefully reduced all-sky
$UBV$ observations obtained during several seasons at Hvar (station~01)
were added to the appropriate magnitude differences for all differential
data sets from the literature.
We also tabulate the mean all-sky values derived from the final reduction
of Fairborn APT and Sejong observations. For comparison,
the mean $V$ magnitude based on the transformation of the Hipparcos
$H_p$ magnitude to Johnson $V$ after \citet{hpvb} is also tabulated for all
comparison stars used.
The number (No.) of individual calibrated all-sky observations
on which the mean values are based is also given.}
\label{comp}
\begin{flushleft}
\begin{tabular}{rccrrrcrr}
\hline\hline\noalign{\smallskip}
Star \ \ \ \ \ &  HD   & $V$&$(B-V)$&$(U-B)$&No.&Station \\
\noalign{\smallskip}\hline\noalign{\smallskip}
BD$+33^\circ4062$&  --  &9\m402\p0.011&   0\m487&   0\m018&  9&01\\
\noalign{\smallskip}\hline\noalign{\smallskip}
BD$+34^\circ4180$&198692&6\m647\p0.011&   1\m018&   0\m802& 33&01\\
BD$+34^\circ4180$&198692&6\m637\p0.007&   1\m027&   0\m832& 11&16\\
BD$+34^\circ4180$&198692&6\m645\p0.008&  --     &   --    &146&61\\
BD$+34^\circ4180$&198692&6\m633\p0.011&   1\m029&   0\m811& 56&78\\
\noalign{\smallskip}\hline\noalign{\smallskip}
         HR 7996 &198820&6\m421\p0.007&$-$0\m132&$-$0\m618&  8&01\\
         HR 7996 &198820&6\m424\p0.007&  --     &   --    &214&61\\
\noalign{\smallskip}\hline\noalign{\smallskip}
BD$+34^\circ4196$&199007&7\m953\p0.014&$-$0\m059&$-$0\m310& 72&01\\
BD$+34^\circ4196$&199007&7\m950\p0.007&$-$0\m059&$-$0\m319& 11&16\\
BD$+34^\circ4196$&199007&7\m959\p0.012&  --     &   --    &138&61\\
BD$+34^\circ4196$&199007&7\m955\p0.014&$-$0\m060&$-$0\m310& 99&78\\
\noalign{\smallskip}\hline\noalign{\smallskip}
BD$+37^\circ4235$&202349&7\m354\p0.009&$-$0\m167&$-$0\m971&149&01\\
BD$+37^\circ4235$&202349&7\m356\p0.010&         &         &161&61\\
\noalign{\smallskip}\hline\noalign{\smallskip}
          70 Cyg &204403&5\m307\p0.009&$-$0\m149&$-$0\m654&108&01\\
          70 Cyg &204403&5\m306\p0.007&$-$0\m149&$-$0\m655& 58&16\\
          70 Cyg &204403&5\m307\p0.006&  --     &   --    &200&61\\
          70 Cyg &204403&5\m307\p0.012&$-$0\m149&$-$0\m655&109&78\\
\noalign{\smallskip}\hline\noalign{\smallskip}
\end{tabular}
\end{flushleft}
Stations:\\
1... Hvar; 16... Fairborn T5; 61... Hipparcos; 78... Sejong
\end{table*}

\end{appendix}
\end{document}